\documentclass[aps,prl,preprint,
superscriptaddress]{revtex4}
\usepackage{graphicx}

\begin{document}

\title{Superconductivity in a single C$_{60}$ transistor}

\author{Clemens. B. Winkelmann*}
\affiliation{Institut N\'eel, CNRS and Universit\'e Joseph Fourier, BP 166, F-38042 Grenoble, France}
\affiliation{Phelma, Grenoble INP, BP 257, F-38016 Grenoble, France}

\author{N. Roch}
\affiliation{Institut N\'eel, CNRS and Universit\'e Joseph Fourier, BP 166, F-38042 Grenoble, France}
\author{W. Wernsdorfer}
\affiliation{Institut N\'eel, CNRS and Universit\'e Joseph Fourier, BP 166, F-38042 Grenoble, France}
\author{V. Bouchiat}
\affiliation{Institut N\'eel, CNRS and Universit\'e Joseph Fourier, BP 166, F-38042 Grenoble, France}
\author{F. Balestro}
\affiliation{Institut N\'eel, CNRS and Universit\'e Joseph Fourier, BP 166, F-38042 Grenoble, France}

\date{\today}

\pacs{74.70.Wz  74.50.+r  73.63.Kv 73.63.Rt}

\maketitle

\textbf{Single molecule transistors (SMTs) are currently attracting enormous attention as possible quantum information processing devices \cite{Loss:01,Romeike:06,Heersche:06,Jo:06,Grose:08}. An intrinsic limitation to the prospects of these however is associated to the presence of a small number of quantized conductance channels, each channel having a high access resistance of at best $R_{K}/2=h/2e^{2}$=12.9 k$\Omega$. When the contacting leads become superconducting, these correlations can extend throughout the whole system by the proximity effect and all contact resistance disappears. This not only lifts the resistive limitation of normal state contacts, but further paves a new way to probe electron transport through a single molecule. In this work, we demonstrate the realization of superconducting SMTs involving a single C$_{60}$ fullerene molecule. The last few years have seen gate-controlled Josephson supercurrents induced in the family of low dimensional carbon structures such as flakes of two-dimensional graphene \cite{Heersche:07} and portions of one-dimensional carbon nanotubes \cite{Jarillo:06}. The present study involving a full zero-dimensionnal fullerene completes the picture.}
\bigskip

 The rapid experimental progress in SMTs has revolutionized our understanding of single-spin magnetism \cite{Heersche:06,Jo:06,Grose:08}, nano-electromechanics \cite{Park:00,Pasupathy:05} and Kondo correlations \cite{Park:02,Liang:02,Yu:05,Roch:08}. However, no experimental realization of superconductivity in a SMT has been reported so far. Superconductivity is a central paradigm to electronic phase coherence phenomena and its effects on SMTs transport properties have been extensively investigated by theory \cite{Choi:04,Novotny:05,Lee:08}. Josephson weak links relying on bottom-up quantum dot (QD) structures could recently be experimentally realized using semiconducting nanowires \cite{Doh:05} as well as carbon nanotubes \cite{Jarillo:06,Cleuziou:06,Jorgensen:06}. Superconductivity in a SMT was yet to be observed as it offers new insights into the couplings involved in single molecule junctions. Superconducting correlations have been observed in ungated junctions involving the metallofullerene Gd@C$_{82}$ \cite{Kasumov:05} but the absence of supercurrents have been attributed to the destructive effect of the local magnetic moment carried by the Gd atom on the superconducting phase coherence. In this Letter, we demonstrate the first experimental realization of superconducting single fullerene molecular transistors, in a broad range of coupling strengths to the contacts, covering both the weakly coupled sequential tunneling regime, as well as the resonant tunneling regime.

SMTs are prepared following the electromigration technique \cite{Park:99}, further optimized as presented in refs. \cite{Park:02,Liang:02,Yu:05,Roch:08}. By extending these advances to measurements at dilution refrigeration temperatures, precise electronic spectroscopies and measurements with superconducting electrodes can be undertaken. The superconducting contacts to the molecular QD are either made of aluminum, or of gold in which superconductivity is induced by the close vicinity of an aluminum capping layer (see Methods for details). After electromigration performed at 4 K, stable devices exhibiting QD behavior are further cooled and measured at 40 mK, with an effective electron temperature of 80 mK. We show here the data on 4 out of 11 samples that were measured at low temperature and which exhibited Coulomb blockade and/or gate-modulated Kondo-like features.


 \begin{figure}[ht]
 \includegraphics[width=0.9\textwidth]{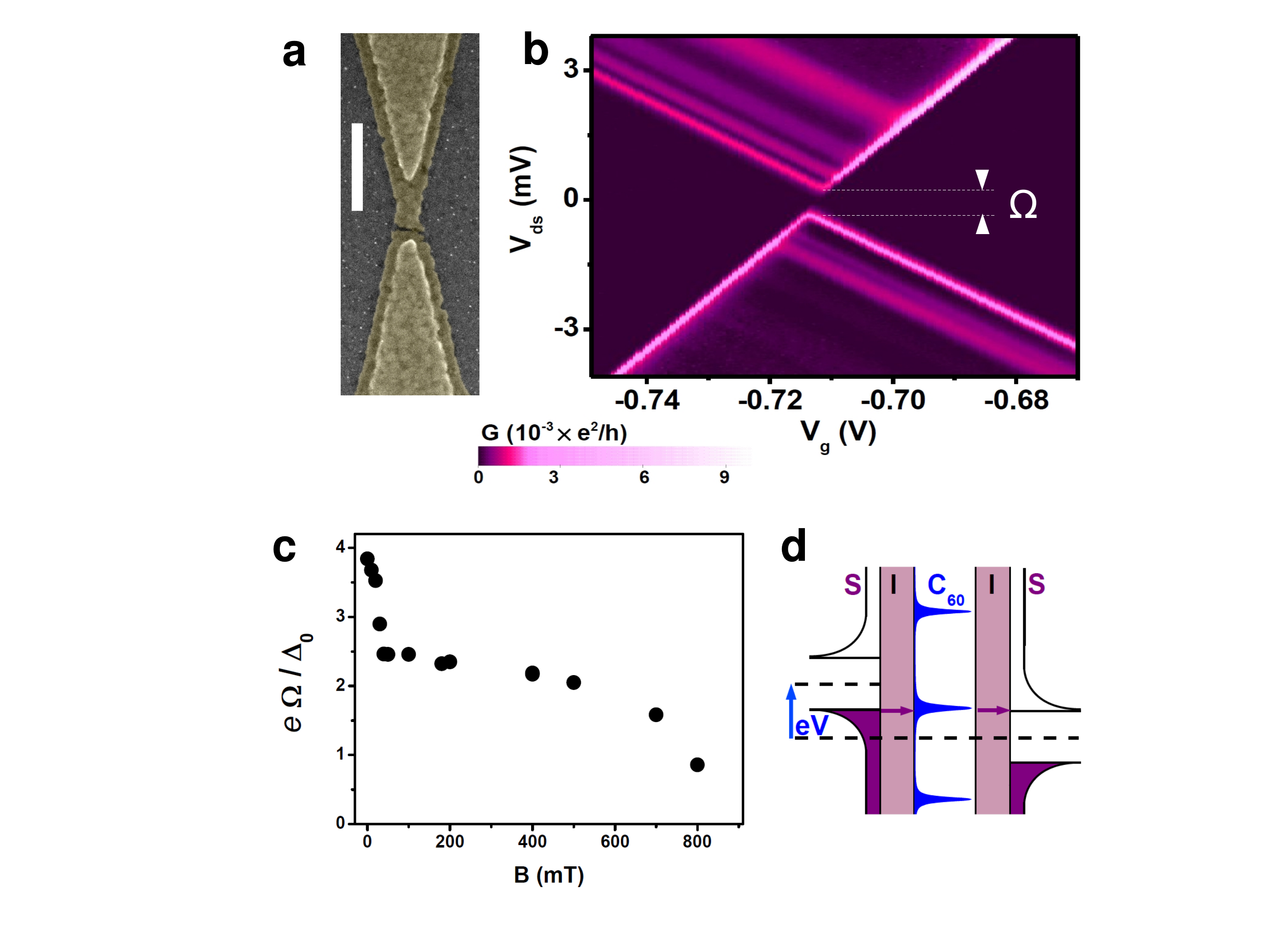}
 \caption{(a) SEM image of an aluminum nanogap obtained by electromigration of a constriction fabricated by angle evaporation. The scale bar is 300 nm. The nanometric gap created by the electromigration process is visible in the center. (b) Differential conductance map of device A as a function of gate and bias voltage, measured at zero magnetic field and T = 35 mK. (c) Extent of the non conducting bias voltage region at degeneracy gate voltage in sample A as a function of magnetic field. (d) Schematics of the level alignment for minimal bias voltage conditions (V$_{ds} = \pm2\Delta_0/e$) allowing for finite current flow.}
 \label{weak}
 \end{figure}

The superconducting weak coupling regime is illustrated in Figure \ref{weak}b, in a device (sample A) displaying two Coulomb diamonds (CDs). In the absence of other conducting regions near the Fermi level over the experimentally accessible gate voltage window (V$_{g}$ = -4.5 to +4 V), an addition energy E$_{add}>250\,$meV can be inferred. This brings clear evidence for single molecular quantum dot behavior because a metallic nanoparticle would lead to a much lower addition energy, typically not exceeding some tens of meV \cite{Black:96}. At base temperature and zero magnetic field, the conducting regions of device A do not intersect the Fermi level: they are separated by a spectroscopic source-drain voltage gap $\Omega$ of about $680\,\mu$V near the degeneracy point. 

This gap, reflecting the quasiparticule spectrum of the contacting electrodes, is a typical feature of a nanostructure weakly coupled to superconducting electrodes \cite{Ralph:95}. In such devices, the current is dominated by incoherent sequential tunneling of quasi-particles of energy $|$E$|>\Delta_0$ (where $\Delta_0=177\, \mu$eV is the bulk superconducting gap of aluminum). The extent $\Omega$ of the non-conducting region is very close to 4$\Delta_0/e$, as expected in an S-I-S tunneling structure (Figure \ref{weak}c,d). The source and drain contacts in a SMT usually have a strong geometrical asymmetry in electromigration devices (Figure \ref{weak}a). In nanostructured aluminum, superconductivity can stabilize up to much higher fields than in bulk material, exceeding 3 T in nanoscaled aluminum particles \cite{Black:96}; one therefore may expect very different critical fields in each contact. This is shown in  Figure \ref{weak}c: $\Omega$ rapidly decreases from 4$\Delta_0/e$ to about 2$\Delta_0/e$ below H$_{c}$' = 40 mT, indicative of one lead going normal, whereas superconductivity in the second lead subsists up to H$_{c}\approx0.85\,$T, the junction being therefore of N-QD-S type at intermediate fields.


 \begin{figure}[ht]
 \includegraphics[width=0.9\textwidth]{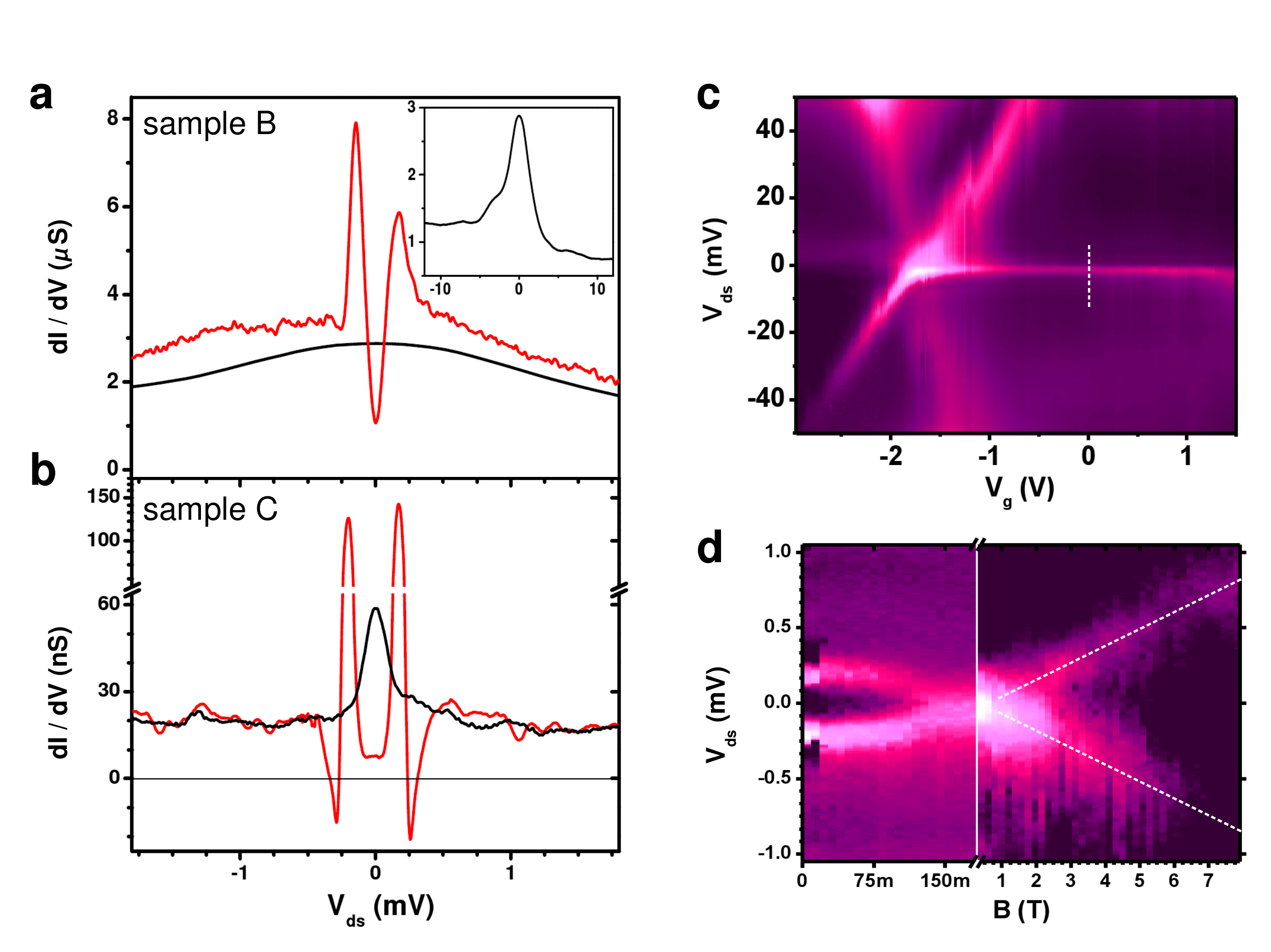}
 \caption{(a) and (b) Differential conductance as a function of bias voltage at constant gate in two samples, at T below 40 mK and in both the normal (black line) and the superconducting state (red line). (a) Sample B has T$_{K} = 14$ K, and superconductivity superimposes on top of the Kondo resonance. The inset shows the normal state data over a larger bias voltage window. (b) Sample C has T$_{K} = 0.7$ K, and the onset of superconductivity suppresses the Kondo resonance. The tunnel coupling of the BCS-density of states to a discrete level leads here to a negative differential conductance close to V$_{ds} = \pm2\Delta_0/e$. Above 65 nS, the vertical scale is logarithmic. (c) Differential conductance plot of sample B in the normal state ($T=40$ mK, $B=400$ mT). The plots shown in (a) are taken along the dotted line (the color code can be read from the vertical scale in (a)). (d) Detailed magnetic field dependence of the low energy differential conductance in sample C. The coherence peaks merge into the Kondo peak at low fields, which is further Zeeman split above $B_{c}\approx0.4\,$T following $E_{Z}=\pm g \Delta S\, \mu_{B}\, (B-B_{c}) $. As $\Delta S=1$, we find here $g=1.84\pm0.2$ along the best linear fit (dotted lines). For better contrast, the differential conductance color code is different in the low and high field regions (the color code can be read from the vertical scale in (b)).}
 \label{Kondo}
 \end{figure}

For increasing coupling to the electrodes (i.e. higher conductance, assuming constant asymmetry and number of conducting channels) co-tunneling events can occur at low bias voltage and enhance the conductance via the Kondo effect at low temperature. The increase of the QD DOS by virtue of the Kondo effect originates from quasiparticles in the contacts lying close to the Fermi level, hopping on and off the dot. The Kondo effect and the suppression of the leads' density of states (DOS) at low energies caused by superconductivity, are therefore rather antagonistic processes. We illustrate the competition between Kondo correlations and superconductivity by two devices, B and C, with Kondo temperatures much larger and lower than  T$_{c}$ respectively. In the normal state of device B, a coulombian addition energy E$_{add}>80\,$meV is observed to coexist with rather strong lead coupling, translating into a large Kondo resonance with T$_{K}=14\,$K (Figure \ref{Kondo}c, details in the Supplementary Information). In the superconducting state of the leads, superconductivity does not suppress the Kondo resonance and the superconducting features such as the coherence peaks at V$_{ds} = \pm2\Delta_0/e$ superimpose on top of it (Figure \ref{Kondo}a). Such a behavior was predicted \cite{Glazman:89,Choi:04} and previously observed \cite{Buitelaar:02,Buizert:07} in other types of QD devices verifying  $T_{K}\gg T_{c}$. The opposite limit $T_{K} <T_c$ is illustrated by the more weakly coupled device C (Figure \ref{Kondo}c), which similarly displayed a Coulomb blockade and Kondo resonance pattern, but with a much lower Kondo temperature T$_{K}=0.70\,$K. In the superconducting state, the Kondo peak at the Fermi level is destroyed \cite{Buitelaar:02,Choi:04}. Interestingly, inside this CD, corresponding to an odd occupation number, i.e. to a local magnetic moment no longer screened by the lead electrons, the differential conductance is not maximum at $V_{ds} = \pm2\Delta/e$. Further, a sign inversion of the differential conductance is observed, which is most prominent close to the degeneracy points. This negative differential conductance, as already reported in larger superconducting QD systems \cite{Ralph:95,Eichler:07,Doh:08}, results from Fermi's golden rule applied to the superconducting leads' DOS, which are convoluted with the delta function-like DOS of the QD. The gradual reduction of the superconducting gap under applied magnetic field reestablishes the Kondo resonance, which is further Zeeman split above about 400 mT (Figure \ref{Kondo}d).


 \begin{figure}[ht]
 \includegraphics[width=0.9\textwidth]{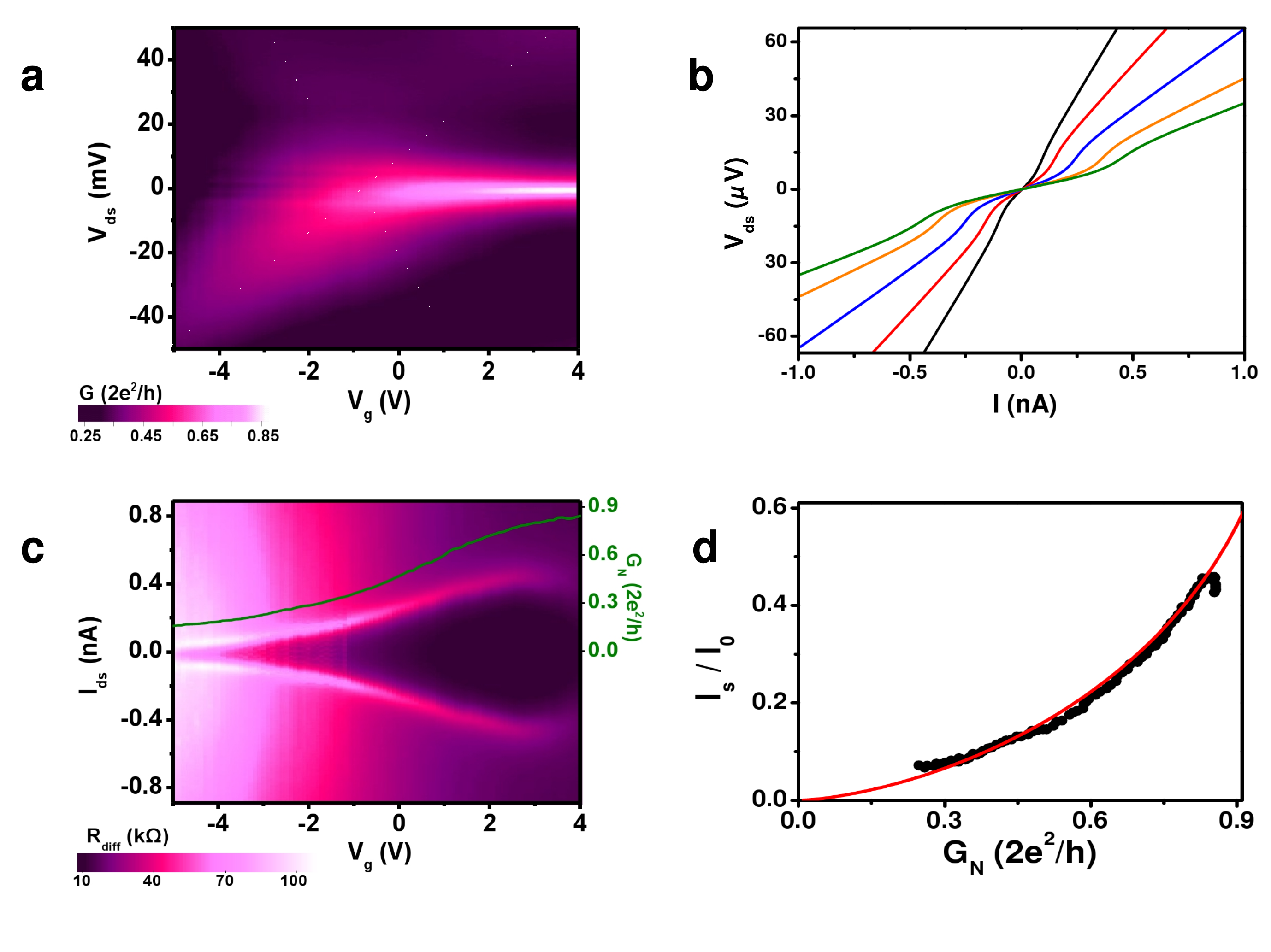}
 \caption{(a) Differential conductance map of device D as a function of gate and bias voltage in the normal state (B = 20 mT). The dotted lines emphasize the CD edges. (b) $V_{ds}(I)$ in the superconducting state for V$_{g}$ = -3, -1.5, 0, 1.5, 3 V from black to green. No hysteresis is observed here, and the differential resistance does not go to zero below the switching current, but to some saturation value in the 5-10$\,$k$\Omega$ range. (c) Differential \textit{resistance} map in the superconducting state (B = 0 T) for small bias currents modulated by I$_{mod} = 18\,$pA at f$_{0}=86\,$Hz. The green line is the zero bias dirrerential conductance in the normal state. (d) Switching current $I_{s}$ normalized to $I_{0}=1.01\,$nA as a function of normal state conductance (bullets), and fit to equation (2) (line). The only adjustable parameter is $I_{0}$. The data points deviating from the fit at high conductance correspond to measurements taken at $V_{g}\approx4$ V, where a significant tunnel leakage current from the gate electrode contaminated the measurements.}
 \label{supra}
 \end{figure}

Moving on to high coupling strengths $\Gamma$, we show that Josephson supercurrent can be observed in SMTs. In equilibrium, dissipationless transport accross a S-QD-S structure is achieved by co-tunneling of Cooper pairs. The Josephson current amplitude $I_c$ therefore scales with $\Gamma^{2}$ and is experimentally only accessible for relatively strong coupling to the leads. Device D illustrates the molecular junction behavior in this regime. In the normal state (Figure \ref{supra}a), the differential conductance map exhibits a Kondo resonance for positive gate voltages on a ridge delimited by only faintly visible CD edges. In this regime the electrostatic on-site repulsion is largely overcome by the strong coupling to the leads. The maximum differential conductance $G_{N}\sim0.85\times (2e^2/h)$ observed here approaches the unitary limit, while the width of the Kondo resonance is associated to a large T$_{K}=80$ K, as already reported in normal SMTs in the past \cite{Yu:04,Yu:05}. In voltage biased measurements of the superconducting state, the differential conductance around the Fermi level displays the usual coherence peaks at $V_{ds} = \pm2\Delta/e$ with $\Delta\approx120\,\mu$eV (not shown, similar to figure \ref{Kondo}a). For supercurrent measurements however, a current bias is better adapted. Figure \ref{supra}b shows $V_{ds}(I)$ measurements for different gate voltages, showing the gate dependence of the critical current. In further detail, Figure \ref{supra}c shows  a $\partial V/\partial I$ differential \textit{resistance} map for small bias currents. The ridge visible here defines a switching current $I_{s}$ such that $\partial^{2} V/\partial I^{2} \mid _{I_{s}}=0$. For $\vert I \vert < I_{s}$ the differential resistance is strongly reduced below $R_j$. We interpret the residual resistance below $I_{s}$ as related to phase diffusion of the Josephson state along the 'tilted washboard' potential [\textbf{cite}]. No hysteresis is observed here which indicates that the Josephson junction is in, or close to, the overdamped limit. The strong reduction of the switching current with respect to  the ideal critical current value  $I_{0}(R_{j})=\Delta/eR_{j}$ (equal to 6 to 20 nA here, depending on the gate voltage) is generic to small Josephson junctions, and is understood as an effect of finite temperature and environmental coupling (e.g. to phonons \cite{Novotny:05}). In particular the critical current dependence on the normal state conductance $G_{N}$ is expected \cite{Joyez:94,Jarillo:06} to follow 
\begin{equation}
I_{c}/I_{0}= (1-\sqrt{1-G_{N} h /2e^2})^{3/2}
\end{equation}
in an unshunted junction hosting a single spin-degenerate conductance channel. Experimentally, this 3/2 power-law dependence was observed to hold in metallic superconducting single electron transistors \cite{Joyez:94} and in carbon nanotube junctions \cite{Jarillo:06}. The fit of the $I_{s}(G_{N})$ data shown in figure \ref{supra}d to (1) yields $I_{0}=1.01\,$nA. It is remarkable that whereas the justification of (1) strictly holds for the underdamped regime \cite{Joyez:94}, a very good agreement to experiment is still found in our device D.

In conclusion, we have shown a full characterization of C$_{60}$ SMTs connected to superconducting leads, covering all coupling regimes. The data demonstrate the coexistence and competition of the effects of Coulomb repulsion, Kondo correlations and superconductivity in SMTs for various coupling strengths to the leads. In the most strongly coupled devices, the unitary conductance limit is approached in the normal state and for the first time a gate modulated superconducting current flowing through a single C$_{60}$ molecule is observed. The observed transport properties are in good agreement with the expected behaviour of a quantum dot connected to superconducting leads. In contrast with previous experiments involving carbon nanotube junctions \cite{Jarillo:06} the energy spectrum is fixed by the molecular edifice itself and not by the nanoengineered portion of nanotube linking the contacting electrode. This experiment paves the way for further studies targetting at phase-sensitive measurements and the interplay of superconductivity with magnetically active molecules such as endofullerenes \cite{Grose:08}.

\bigskip
\textbf{ METHODS }
{\small

The starting material for the devices are thin metallic wires with a local constriction about 100 nm wide and 20 nm thick, defined by e-beam lithography and double angle evaporation through a suspended mask, on top of a local Al/Al$_{2}$O$_{3}$ backgate (Figure \ref{weak}a), as described in \cite{Roch:08}. C$_{60}$ diluted in toluene is drop-casted on the thoroughly cleaned devices which are then immediately introduced into the sample mount thermally anchored to the mixing chamber of a dilution refrigerator. At liquid helium temperatures, the bias voltage across the wires is ramped up until the electromigration process sets on. The device resistance, which has an initial value of about 100 $\Omega$, increases, often stepwise, indicating the breaking up of individual 1D atomic conduction channels. As soon as the device resistance exceeds 12.9 k$\Omega$, the bias voltage is cancelled within microseconds. After electromigration, the junctions are further cooled down to sub-0.1 K temperatures. About 50 \% of all electromigrated junctions display a tunneling transport behavior with a low bias resistance $R_{K}<R_{j}<1\,$G$\Omega$. About 10 \% among these further show a gate dependent signal, indicative of QD behavior. High bandwidth control of the electromigration process is known to strongly increase the yield for obtaining narrow tunneling junctions. Two types of superconducting electromigration junctions have been investigated : (i) all aluminum electromigration junctions and (ii) proximity superconducting Al/Au junctions. In the later, the tunnel junction is opened through a short ($L<150 $nm) and thin gold wire connected to nearby larger superconducting aluminum source and drain pads (for details see Supp. Info.).
The method used here for nanogap formation requires to get a low impedance environment in order to keep the bandwith during electromigration high enough during conductance monitoring. This forbids the implementation of chip-resistors therefore preventing a full control of the electromagnetic environment.}

\bigskip

\textbf{Acknowledgments}
We would like to thank E. Eyraud, Th. Fournier and D. Lepoittevin for technical assistance, T. Martin, M. Lee, D. Feinberg, M. Deshmukh and H. Courtois for stimulating discussions. This project was funded by the French ANR/PNANO Molspintronics and ANR/JC NEMESIS.

\textbf{Author Information } The authors declare no competing financial interests. Correspondence should be addressed to C.B.W. (clemens.winkelmann@grenoble.cnrs.fr).

\end{document}